\documentclass[aps,pre,showpacs,amsmath,amssymb,amsfonts,superscriptaddress,lengthcheck,twocolumn,longbibliography]{revtex4-1}
\usepackage{graphicx}
\usepackage{subfigure}
\usepackage{amsthm}
\usepackage{verbatim}
\usepackage{dcolumn}
\usepackage{bm}
\usepackage{epsf}
\usepackage{color}
\usepackage[colorlinks=true,citecolor=blue,linkcolor=blue,urlcolor=blue]{hyperref}%
\usepackage{xcolor}
\usepackage{dsfont}
\usepackage{tikz}
\usetikzlibrary{matrix,arrows}
\usepackage{bbold}
\usepackage{float}

\newcommand{\tr}[1]{\mathrm{tr}\left\{#1\right\}}
\newcommand{\ptr}[2]{\mathrm{tr_{#1}}\left\{#2\right\}}

\newcommand{\etal}{\textit{et al. }}

\newcommand{\bla}{bla\\bla\\bla\\bla\\bla}

\newcommand{\mrm}[1]{\mathrm{#1}}

\usepackage{soul}

\begin{document}

\title{Non-thermal quantum engine in transmon qubits}

\author{Cleverson Cherubim}
 \email{cleverson.cherubim@usp.br}
\author{Frederico Brito}%
 \email{fbb@ifsc.usp.br}
\affiliation{Instituto de F\'isica de S\~ao Carlos, Universidade de S\~ao Paulo, C.P. 369, 13560-970, S\~ao Carlos, SP, Brazil}%
\author{Sebastian Deffner}%
 \email{deffner@umbc.edu}
\affiliation{Department of Physics, University of Maryland Baltimore County, Baltimore, MD 21250, USA}

\date{\today}

\begin{abstract}
The design and implementation of quantum technologies necessitates the understanding of thermodynamic processes in the quantum domain. In stark contrast to macroscopic thermodynamics, at the quantum scale processes generically operate far from equilibrium and are governed by fluctuations. Thus, experimental insight and empirical findings are indispensable in developing a comprehensive framework. To this end, we theoretically propose an experimentally realistic quantum engine that uses transmon qubits as working substance. We solve the dynamics analytically and calculate its efficiency.
\end{abstract}

\maketitle


\section{\label{sec:level1}Introduction}

Recent advances in nano and quantum technology will necessitate the development of a comprehensive framework for \emph{quantum thermodynamics} 
 \cite{Gemmer2004}. In particular, it will be crucial to investigate whether and how the laws of thermodynamics apply to small systems, whose dynamics are governed by fluctuations and which generically operate far from thermal equilibrium. In addition, it has already been recognized that at the nanoscale many standard assumptions of classical statistical mechanics and thermodynamics are no longer justified and even in equilibrium quantum subsystems are generically not well-described by a Maxwell-Boltzmann distribution, or rather a Gibbs state \cite{Gelin2009}. Thus, the formulation of the statements of quantum thermodynamics have to be carefully re-formulated to account for potential quantum effects in, for instance, the efficiency of heat engines \cite{Scully862,PhysRevE.92.042126,Deffner2018Entropy,Cakmak2019}.

In good old thermodynamic tradition, however, this conceptual work needs to be guided by experimental insight and empirical findings. To this end, a cornerstone of quantum thermodynamics has been the description of the working principles of quantum heat engines \cite{PhysRevX.7.031044,0295-5075-88-5-50003,PhysRevE.86.051105,0295-5075-106-2-20001,PhysRevLett.112.030602,2015NatSR512953H,1367-2630-18-8-083012,PhysRevE.93.052120,PhysRevB.96.104304,PhysRevE.90.012119,PhysRevE.88.032103}.

However, to date it is not unambiguously clear whether quantum features can always be exploited to outperform classical engines, since to describe the thermodynamics of non-thermal states one needs to consider different perspectives---different than the one established for equilibrium thermodynamics. For instance, it has been shown that the Carnot efficiency cannot be beaten \cite{PhysRevE.92.042126,Gardas2016SciRep} if one accounts for the energy necessary to maintain the non-thermal stationary state \cite{PhysRevLett.86.3463,doi:10.1143/PTPS.130.29,Horowitz2014,Yuge2013}. However, it has also been argued that Carnot's limit can be overcome, if one carefully separates the ``heat" absorbed from the environment into two different types of energy exchange \cite{PhysRevE.91.032119,gers}: one is associated with a variation in \textit{passive energy} \cite{Pusz1978,0295-5075-67-4-565} which would be the part responsible for changes in entropy, and the other type  is a variation in \textit{ergotropy}, a work-like energy that could be extracted by means of a suitable unitary transformation. On the other hand, it has been shown \cite{PhysRevE.98.042123} that a complete thermodynamic description in terms of \textit{ergotropy} is also not always well suited. Having several perspectives to explain the same phenomenon is a clear indication of the subtleties and challenges faced by quantum thermodynamics, and which can only be settled by the execution of purposefully designed experiments. Therefore, theoretical proposals for feasible and relevant experiments appear instrumental.

In this work we propose an experiment to implement a thermodynamic engine with a transmon qubit as the working substance (WS), which interacts with a non-thermal environment composed by two subsystems, an externally excited cavity (a superconducting transmission line) and a classical heat bath \cite{0957-4484-27-36-364003} with temperature $T$. The WS undergoes a non-conventional cycle (different from Otto, Carnot, etc.) \cite{callen1985thermodynamics} through a succession of non-thermal stationary states obtained by slowly varying its bare energy gap (frequency) and the amplitude of the pumping field applied to the cavity. We calculate the efficiency of this engine for a range of experimentally accessible parameters \cite{0957-4484-27-36-364003, kok2007,hofheinz2008,mallet2009}, obtaining a maximum value of $47\%$, which is comparable with values from the current literature.

\section{\label{describing_system}System description}

\begin{figure}[H]
\centering
\includegraphics[scale=.4]{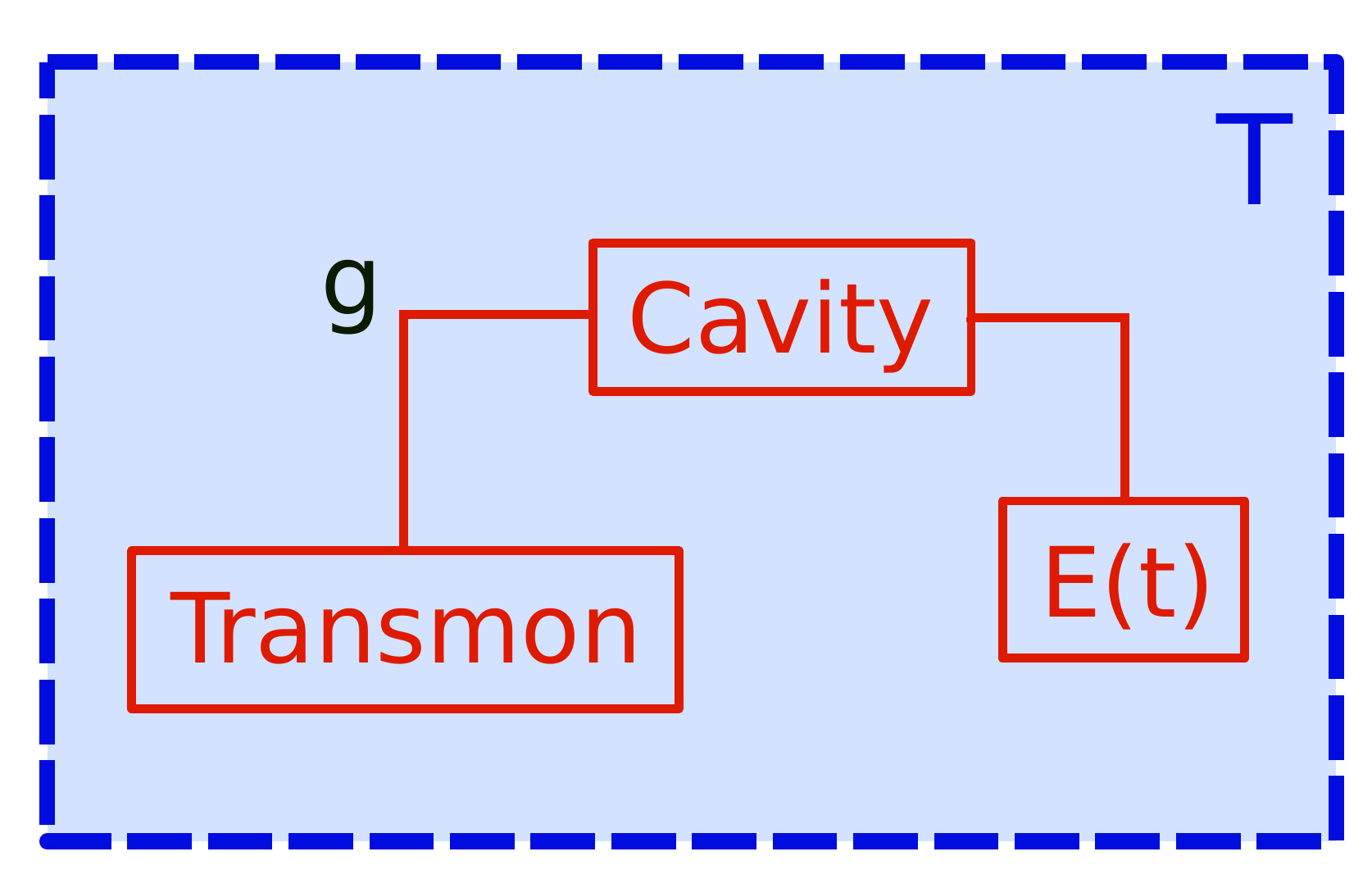}
\centering
\caption{\label{system_sketch} Sketch of the quantum engine with a transmon qubit as working substance interacting with an externally pumped (E(t)) transmission line (cavity). Both systems are embedded in the same cryogenic environment, which plays the role of a standard thermal bath of temperature $T$. Such a setup gives a dynamics of a working substance in the presence of a controllable \emph{non-thermal} environment.	}\label{fig1}
\end{figure}

We consider a multipartite system, comprised of a transmon qubit of tunable frequency $\omega_\mrm{T}$, which interacts with a transmission line (cavity) of natural frequency $\omega_\mrm{CPW}$ with coupling strength $g$. The cavity is pumped by an external field of amplitude $E_d$ and single frequency $\omega$ (see Figure \ref{fig1}). Both systems are in contact with a classical heat bath at temperature $T$. Such a set-up is experimentally realistic and several implementations have already been reported in different contexts \cite{Majer2007,0957-4484-27-36-364003}. Here and in the following, the transmon is used as a working substance (WS) and the (non-standard) ``bath'' is represented by the net effect of the other two systems: the cavity and the cryogenic environment (classical bath). There are two subtleties that must be noted here: (i) the bath ``seen'' by the qubit does not only consist of a classical reservoir at some fixed temperature, but it has an additional component, namely the pumped cavity. By changing the pumping, several cavity states can be realized. Such a feature gives the possibility of making this composed bath \emph{non-thermal} on demand. In addition, (ii), the proposed engine is devised as containing only one bath (cavity + environment), which does not pose any problems considering that it is an out-of-equilibrium bath.

We start our analysis from the Hamiltonian describing a tunable qubit interacting with a single mode pumped cavity through a Jaynes-Cummings interaction
\begin{equation}
\label{hamiltonian_complete_system}
\begin{split}
H(t)&=\frac{\hbar\omega_\mrm{T}}{2}\sigma_z+\hbar\omega_\mrm{CPW} a^\dagger a+g\sigma_x(a+a^\dagger) \\
&+E_d\left(ae^{i\omega t}+a^\dagger e^{-i\omega t}\right),
\end{split}
\end{equation}	
where $\sigma_x$ and $\sigma_z$ are the Pauli matrices, $a^\dag$ and $a$ are the canonical bosonic creation and annihilation operators associated with the cavity excitations, $g$ is the qubit-cavity coupling strength. The last term represents a monochromatic pumping of amplitude $E_d$ and frequency $\omega$ applied to the cavity. The experimental characterization of the qubit-cavity dissipative dynamics emerging from their interaction with the same thermal bath shows that the system's steady state is determined by the master equation~\cite{0957-4484-27-36-364003}
\begin{equation}
\label{complete_master_equation_rotating_reference}
\begin{split}
\dot{\rho}(t)&=-\frac{i}{\hbar}[H_\mrm{RWA},\rho]+K^{-}_\mrm{CPW}\mathcal{D}[a]\rho \\
 &+K^{+}_\mrm{CPW}\mathcal{D}[a^\dagger]\rho+\Gamma^{-}\mathcal{D}[\sigma^{-}]\rho+\Gamma^{+}\mathcal{D}[\sigma^{+}]\rho,
 \end{split}
\end{equation}
with $K^{-}_\mrm{CPW}(K^{+}_\mrm{CPW})$ being the cavity decay (excitation) rate, $\Gamma^{-} (\Gamma^{+})$ the qubit relaxation (excitation) rate and $\mathcal{D}[A]\rho=A\rho A^\dag-1/2(A^\dag A\rho+\rho A^\dag A)$. Please note that these rates satisfy detailed balance for the same bath of temperature $T$, $K^{+}_\mrm{CPW}/K^{-}_\mrm{CPW}=\exp{(-\hbar\omega_\mrm{CPW}/k_\mrm{B}T)}$ and $\Gamma^{+}/\Gamma^{-}=\exp{(-\hbar\omega_\mrm{T}/k_\mrm{B}T)}$. The Hamiltonian part
\begin{equation}
\label{hamiltonian_complete_system_rwa}
\begin{split}
H_\mrm{RWA}&=\frac{\hbar}{2}\left(\omega_\mrm{T}-\omega\right)\sigma_z+\hbar\left(\omega_\mrm{CPW}-\omega\right) a^\dagger a  \\
&+g(\sigma_{+}a+\sigma_{-}a^\dagger)+E_d(a+a^\dagger), \\
\end{split}
\end{equation}
is the system Hamiltonian in the rotating wave approximation (RWA) \cite{scully_zubairy_1997}, with $\sigma_{+}$($\sigma_{-}$) being the spin ladder operators.

Since we are interested in the observed dynamics of the WS, it is necessary to find the qubit's reduced density matrix $\rho_\mrm{T}(t)\equiv \ptr{a}{\rho(t)}$, where $\ptr{a}{\cdot}$ represents the partial trace on the cavity's degrees of freedom. The system state is in a qubit-cavity product state, i.e., $\rho(t)\approx\rho_\mrm{T}(t)\otimes\rho_\mrm{C}(t)$, which emerges in the effective qubit-cavity weak coupling regime due to decoherence into the global environment. In addition, the cavity's stationary state $\rho_\mrm{C}(t)$ is assumed to be mainly determined by the external pumping, which can be easily found for situations of strong pumping and/or {weak coupling strength $g$}. This closely resembles a situation, in which the cavity acts as a work source of effectively infinite inertia \cite{PhysRevX.3.041003}. Thus, changing the state of the qubit does not affect the state of the cavity, but it is still susceptible to the applied field and the cryogenic bath, and we have
\begin{equation}
\langle a\rangle = {\langle a^{\dagger}\rangle}^{*}=\frac{E_d}{\hbar\left[i\kappa_\mrm{CPW}/2-(\omega_\mrm{CPW}-\omega)\right]},\label{coherent}
\end{equation}
{where we defined $K^{-}_\mrm{CPW}=\kappa_\mrm{CPW}$}. Hence, the reduced master equation \eqref{complete_master_equation_rotating_reference} can be written as 
\begin{equation}
\label{transmon_master_equation_rotating_reference}
\dot{\rho}_\mrm{T}(t)=-\frac{i}{\hbar}[\tilde{H}_\mrm{T,RWA},\rho_\mrm{T}]+\Gamma^{-}\mathcal{D}[\sigma^{-}]\rho_\mrm{T}+\Gamma^{+}\mathcal{D}[\sigma^{+}]\rho_\mrm{T},
\end{equation}
with
\begin{equation}
\label{transmon_hamiltonian}
\tilde{H}_\mrm{T,RWA}=\frac{\hbar}{2}(\omega_\mrm{T}-\omega)\sigma_z+g\left[\langle a\rangle\sigma_{+}+\langle a^\dagger\rangle\sigma_{-}\right].
\end{equation}

Please note that the effective qubit Hamiltonian carries information about the interaction with the cavity through $\langle a\rangle$ and $\langle a^\dagger\rangle$, which are dependent on the cavity state.

\section{Non-equilibrium thermodynamics}
\subsection{Non-thermal equilibrium states}

The only processes that are fully describable by means of conventional thermodynamics are infinitely slow successions of equilibrium states. For the operating principles of heat engines, the second law states that the maximum attainable efficiency of a thermal engine operating between two heat baths is limited by Carnot's efficiency. 

An extension of this standard description is considering infinitely slow successions along \emph{non-Gibbsian}, but stationary states \cite{doi:10.1143/PTPS.130.29,PhysRevLett.86.3463,Sasa2006,PhysRevE.92.042126,Gardas2016SciRep}. In the present case, namely a heat engine with transmon qubit as working substance, non-Gibbsianity is induced by the external excitation applied as a driving field to the cavity. We will see in the following, however, that identifying the thermodynamic work is subtle -- and that the energy exchange can exhibit heat-like character, which is crucial when computing the entropy variation during the engine operation.

The stationary state can be found by solving the master equation Equation~\eqref{transmon_master_equation_rotating_reference}, and is written as
\begin{equation}
\label{stad_state}
\rho_\mrm{T}^{ss}=\begin{pmatrix}
    \rho_\mrm{T}^{ee}       & \rho_\mrm{T}^{eg}\\
    \rho_\mrm{T}^{ge}       & \rho_\mrm{T}^{gg}
\end{pmatrix}
\end{equation}
where the matrix elements can be computed explicitly and are summarized in Appendix \ref{sec:app}.

We observe that for the case of effective qubit-cavity ultra-weak coupling, i.e., $\hbar\omega_\mrm{T}\gg g E_d/\left|i\,\hbar\kappa_\mrm{CPW}/2-\hbar(\omega_\mrm{CPW}-\omega)\right|$, as expected, the obtained non-thermal state asymptotically approaches thermal equilibrium, namely $|\rho_\mrm{T}^{eg}|=|\rho_\mrm{T}^{ge}|\approx0$ and $\rho_\mrm{T}^{ee}/\rho_\mrm{T}^{gg}\approx \exp{(-\beta\hbar\omega_\mrm{T})}$. In addition, as also expected, in the high temperature limit $\hbar\omega_\mrm{T}/{kT}\ll1$ the qubit stationary state becomes the thermal, maximally mixed state, given that the cavity is not strongly pumped.

\subsection{The Cycle}

In equilibrium thermodynamics cycles are constructed by following a closed path on a surface obtained by the equation of state \cite{callen1985thermodynamics}, which characterizes possible equilibrium states for a given set of macroscopic variables. This procedure can be generalized in the context of steady state thermodynamics, where an equation of state is also constructed.

\begin{figure}[H]
\includegraphics[scale=1.0]{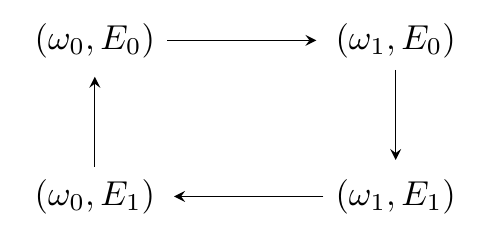}
\centering
\caption{\label{cycle_figure} Sketch of the thermodynamic cycle obtained by varying the tunable parameters $\omega_\mrm{T}$ and $E_d$. Each one of the strokes are obtained by keeping one of the variables constant while quasi-statically varying the other one.}
\end{figure}

For the present purposes, we use the steady state (\ref{stad_state}) to devise a cycle for our heat engine. The equation of state in our case is represented by the stationary state's von Neumann entropy {$S(\omega_\mrm{T},E_d)=-\tr{\rho_\mrm{T}^{ss}\ln{\rho_\mrm{T}^{ss}}}$}, which is fully determined by the pair of controllable variables $\omega_\mrm{T}$, the transmon's frequency, and $E_d$, amplitude field of the pumping applied to the cavity. In order to implement the cycle, the stationary state is slowly varied (quasi-static) \footnote{The timescale for which the changes made can be considered slow is such that the conditions imposed to the system state are satisfied, namely the state is a product state and the cavity steady state is a coherent state with Equation~(\ref{coherent})} by changing the ``knobs'' $(\omega_\mrm{T}$,$E_d)$. It is composed of four strokes where we keep one of the two controllable variables constant and vary the other one, for example, at the first stroke we keep $E_d=E_0$ and vary $\omega_\mrm{T}$ from $\omega_0$ to $\omega_1$. The complete cycle is sketched in Figure~\ref{cycle_figure}.

Since we are interested in analyzing the engine as a function of its parameters of operation, we simulated several cycles with boundary values $(\omega_1,E_1)$, which will range from the minimum value $(\omega_0,E_0)$ to the maximum one $(\omega_{1,\mathrm{max}},E_{1,\mathrm{max}})$. The corresponding cycles lie on the von Neumann entropy surface depicted in Figure~\ref{entropies_plot}. {In Appendix \ref{sec:app} plots of the stationary state's population and quantum coherence $\rho_\mrm{T}^{ee}$ and $|\rho_\mrm{T}^{eg}|$ as a function of $(\omega_\mrm{T},E_d)$ are shown}. There we can observed clearly that the WS exhibits quantum coherence and population changes during its operation.
\begin{figure}[H]
\includegraphics[scale=.7]{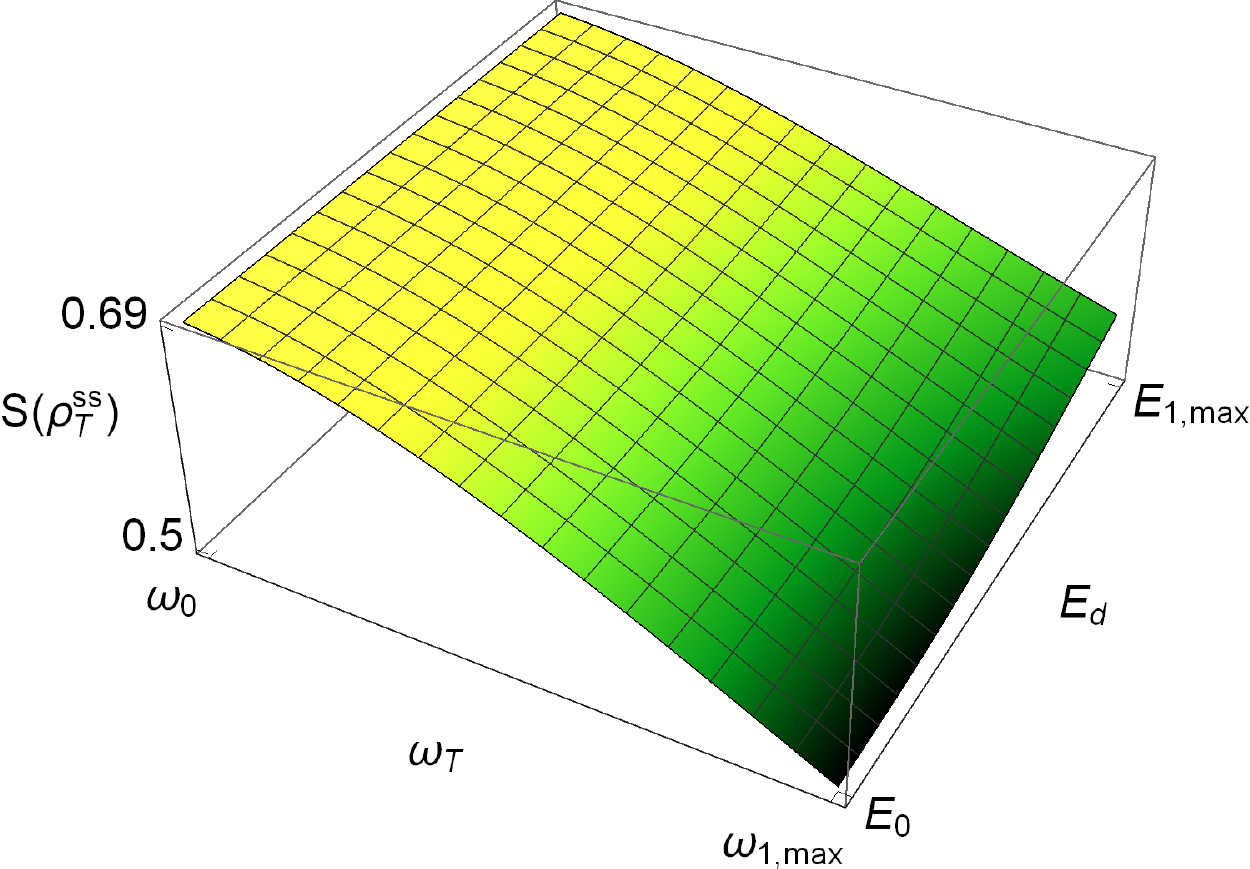}
\centering
\caption{\label{entropies_plot} Stationary state's von Neumann entropy in the regime of operation of the thermal engine. Any thermodynamic cycle must be contained on this surface.}
\end{figure}

Finally, it is worth emphasizing that in the present analysis all parameters were chosen from an \emph{experimentally accessible} regime \cite{0957-4484-27-36-364003, kok2007,hofheinz2008,mallet2009}, under the validity of the approximation of weak-coupling interaction between transmon and cavity. The parameters are collected in Table~\ref{tab_parameters}.
\begin{table}[H]
  \begin{center}
    \begin{tabular}{l|c} 
    \textbf{Parameter} & \textbf{Value} \\
    \hline
      $\omega_\mrm{CPW}/2\pi$ & 4.94 GHz \\
      $\omega/2\pi$ & 4.94 GHz \\
      $g/{2\pi\hbar}$ & 120 MHz \\
      $T$ & 30 mK \\
      $\Gamma/2\pi$ & 2 MHz \\
      $\kappa_{{\rm CPW}}/2\pi$ & 1 MHz \\
      $\omega_0 /2\pi$ & 100 MHz \\
      $\omega_{1,\mathrm{max}} /2\pi$ & 1000 MHz \\
      $E_0 /{2\pi\hbar}$ & 0.2 MHz \\
      $E_{1,\mathrm{max}} /{2\pi\hbar}$ & 2 MHz \\
    \end{tabular}
    \caption{\label{tab_parameters}Engine parameters used in the present analysis.}
  \end{center}
\end{table}

\section{ Work, heat and efficiency}

The first law of thermodynamics, $\Delta E(t)=W(t)+Q(t)$, states that a variation of the internal energy along a thermodynamic process can be divided into two different parts, work $W(t)$ and heat $Q(t)$, where for Lindblad dynamics we have \cite{0305-4470-12-5-007,PhysRevE.92.042126},
\begin{equation}
\label{wq}
\begin{split}
        W(t)&=\int_{0}^{t} \tr{\rho(t')\dot{H}(t')dt'},\\
        Q(t)&=\int_{0}^{t} \tr{\dot{\rho}(t')H(t')dt'}.
\end{split}
\end{equation}
Typically, work is understood as a controllable energy exchange, which can be used for something useful, while heat cannot be controlled, emerging from the unavoidable interaction of the engine with its environment. As stated before, there are certain situations  in which it can be shown that part of $Q(t)$ does not cause any entropic variation \cite{gers}. This has led to proposals for the differentiation of two distinct forms of energy  contributions to $Q$: the \textit{passive energy} $\mathcal{Q}(t)$, which is responsible for the variation in entropy, and the variation in \textit{ergotropy} $\Delta\mathcal{W}(t)$ which is a ``work-like energy'' that can be extracted by means of a unitary transformation and consequently would not cause any entropic change. Both terms are defined as,
\begin{equation}
\label{q_separation}
\begin{split}
		\mathcal{Q}(t)&=\int_{0}^{t} \tr{\dot{\pi}(t')H(t')dt'}, \\
		\Delta\mathcal{W}(t)&=\int_{0}^{t} \tr{[\dot{\rho}(t')-\dot{\pi}(t')]H(t')dt'},
\end{split}
\end{equation}
with $\pi(t)$ being the passive state \cite{Pusz1978} associated with the state $\rho(t)$ at time $t$. To calculate the upper bound on the efficiency for systems that exhibit these different ``flavors'' of energy one should replace $Q$ by $\mathcal{Q}$ in statements of the second law, since the \textit{ergotropy} is essentially a mechanical type of energy, and consequently not limited by the second law, resulting in a different upper bound, see also Ref.~\cite{gers}.

Distinguishing these types of energy exchanged with the environment is crucial when one is interested in determining the fundamental upper bounds on the efficiency. However, in the present context,  we are more interested in experimentally relevant statements, i.e., computing the efficiency in terms of what can be measured directly. Thus, we consider the ratio of the extracted work to the total energy acquired from the bath, independent of its type \cite{gers}.

The cycle designed here is such that in each stroke one of the knobs $(\omega_\mrm{T},E_d)$ is kept fixed, while the other one is changed. Recall that the cavity is assumed to be a subpart of the bath seen by the WS, and that its state is modified by $E_d$. Since the WS is always in contact with the environment, one has that heat and work are exchanged in each stroke. Here, such a calculation is done by using Equation~\eqref{wq}, considering the stationary state Equation~(\ref{stad_state}) and the effective WS Hamiltonian Equation~(\ref{transmon_hamiltonian}). Then, for the $i$th stroke, the corresponding $W_i$ and $Q_i$ integrals, representing the work and heat delivered (extracted) to (from) the WS, can be parametrized in terms of the respective knob variation as we can see in Appendix \ref{therm_quant}. These quantities are obtained using the WS effective Hamiltonian $\tilde{H}_\mrm{T,RWA}$, which already takes into account the interaction with the external bath and pumped cavity.

Once these quantities are determined, we can calculate the efficiency $\eta$ of this engine, defined by
\begin{equation}
\label{eff}
\eta=-\frac{\sum_{i=1}^{4}{W_i}}{Q_{+}},
\end{equation}
with the delivered heat to the WS in a complete cycle being given by $Q_{+}=\sum_{i=1}^{4}{Q_{+}^i}$, with $Q_{+}^i$ the given heat (only positive contributions inside the stroke) during the $i_{\mathrm{th}}$ stroke (see Appendix \ref{therm_quant}). Therefore, this efficiency represents the amount of work extracted from the engine through the use of the delivered heat to the WS. 

\begin{figure}[H]
\includegraphics[scale=.56]{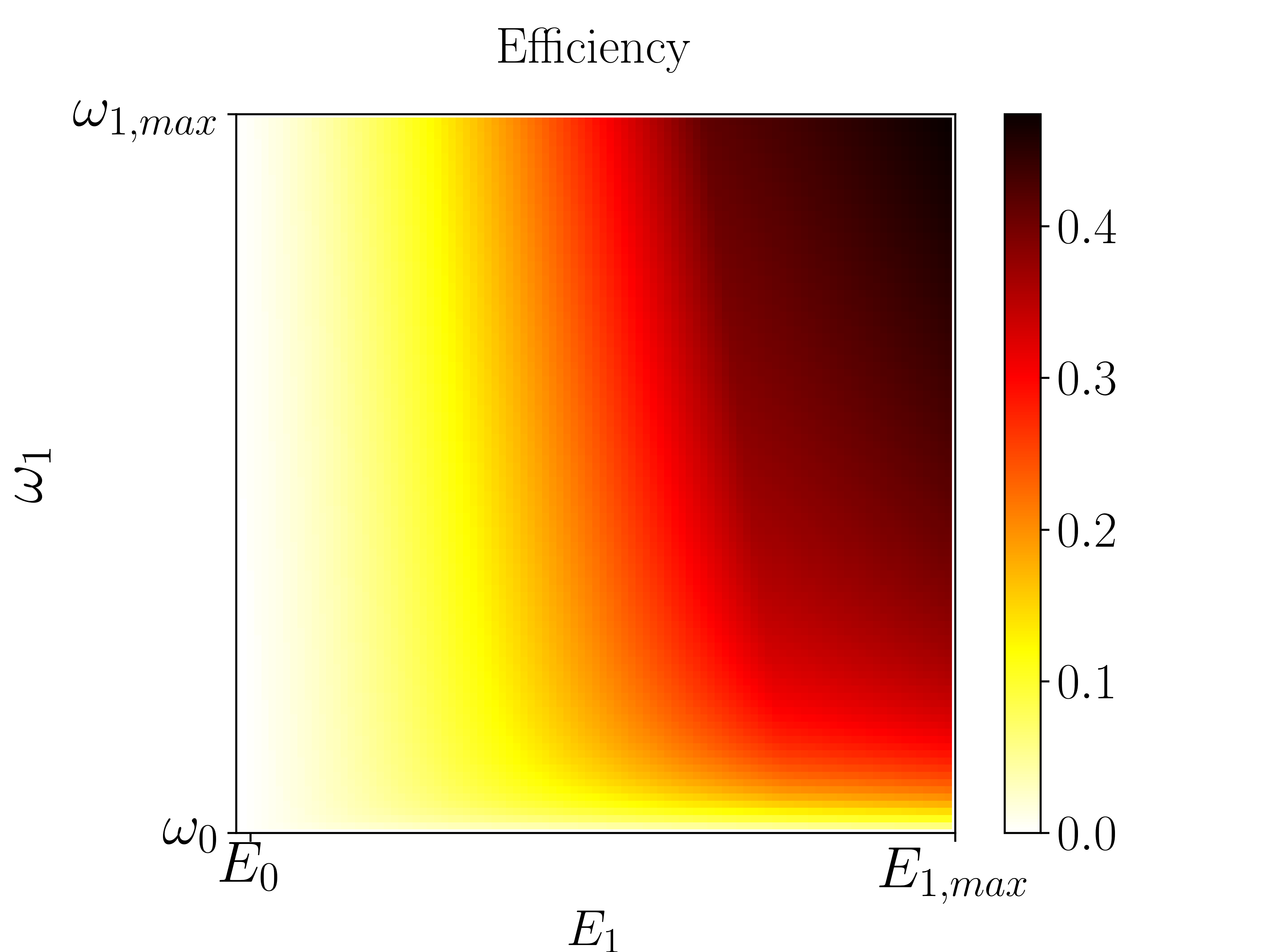}
\centering
\caption{\label{efficiency_plot} Efficiency $\eta$ as a function of the upper values $(\omega_1,E_1)$ for the cycle depicted in Figure~\ref{cycle_figure}. The observed highest efficiency of about $47\%$ was attained when $(\omega_1,E_1)=(\omega_{1,\mathrm{max}},E_{1,\mathrm{max}})$, with $\omega_{1,\mathrm{max}} /2\pi=1000$ MHz and $E_{1,\mathrm{max}} /{2\pi\hbar}=2$ MHz.}
\end{figure}

Figure \ref{efficiency_plot} shows the engine efficiency $\eta$ attained in the execution of the strokes as a function of the boundary values $(\omega_1,E_{1})$, as depicted in Figure \ref{cycle_figure}. Please note that $(\omega_1,E_{1})$ sweeps the entire spectrum of the tunable parameters $(\omega_{\mrm{T}},E_d)$, going from $(\omega_0,E_{0})$ to $(\omega_{1,\mathrm{max}},E_{1,\mathrm{max}})$ where we find the maximal efficiency. It is worth mentioning here that the highest value of the efficiency is dependent on the chosen regime of parameters, which in our case is based on experimentally attainable values~\cite{0957-4484-27-36-364003, kok2007,hofheinz2008,mallet2009}. As usual, in order to extract the predicted work, one has to couple our engine to another system. We~envision  using the experimental setup of Ref. \cite{0957-4484-27-36-364003}, where a mechanical nanoresonator is present and weakly driven by the transmon. Thus, under such a configuration, by following the nanoresonator's state \footnote{recall that we have assumed infinite inertia, i.e., the transmon is not capable of changing the cavity's state. In situations where such an assumption does not hold, one has to take into account the possibility of having the transmon doing work on the cavity}, one can determine the amount of energy transferred in the form of work. In addition, by observing the transmon's state, one can obtain the amount of heat given by the non-standard bath, providing a full characterization of our~engine.
\section{Conclusion and final remarks}

Theoretical research of small heat engines in the quantum domain is common place in quantum thermodynamics \cite{0305-4470-12-5-007,article,PhysRevLett.93.140403,PhysRevE.76.031105,PhysRevLett.105.130401,2014NatSR...4E3949C,PhysRevX.5.031044,PhysRevLett.109.203006,PhysRevLett.112.150602,Dawkins2018}. In the present work, we have devised a transmon-based heat engine using an experimentally realistic regime of parameters reaching a maximal efficiency of $47\%$, which turns out to be a reasonable value when compared with the state of the art in quantum heat engines. One of the most recent experiments in quantum heat engine was implemented by Peterson \etal \cite{2018arXiv180306021P} using a spin $-1/2$ system and nuclear resonance techniques, performing an Otto cycle with efficiency in excess of $42\%$ at maximum power. It is important to stress that implementing small heat engines constitutes a hard task, even when dealing with classical systems. Indeed, a representative example is the single ion confined in a linear Paul trap with a tapered geometry, which was used to implement a Stirling engine \cite{Rob} with efficiency of only $0.28\%$. {Additional research is being carried out concerning the behavior of this engine influenced by the presence of coherence and the dimension of the WS}. By devising this theoretical protocol for the implementation of a quantum engine, we hope to help the community, and in particular experimentalists, in the formidable task to design and implement quantum thermodynamic systems and to consolidate the concepts of this new exiting field of research.

\begin{acknowledgments}
We thank F. Rouxinol and V. F. Teizen for valuable discussions. C.C. would like to thank the hospitality of UMBC, where most of this research was conducted. C.C. and F.B. acknowledge financial support in part by the Coordena\c{c}\~ao de Aperfei\c{c}oamento de Pessoal de N\'ivel Superior - Brasil (CAPES) - Finance Code 001. During his stay at UMBC, C.C. was supported by the CAPES scholarship PDSE/process No. 88881.132982/2016-01. F.B. is also supported by the Brazilian National Institute for Science and Technology of Quantum Information (INCT-IQ) under Grant No. 465469/2014-0/CNPq. S.D. acknowledges support from the U.S. National Science Foundation under Grant No. CHE-1648973.
\end{acknowledgments}
\newpage
\appendix

\onecolumngrid
\section{Non-thermal equilibrium states \label{sec:app}}

\begin{figure}[h]
\centering     
\subfigure[Excited state's population]{\label{fig:b}\includegraphics[width=80mm]{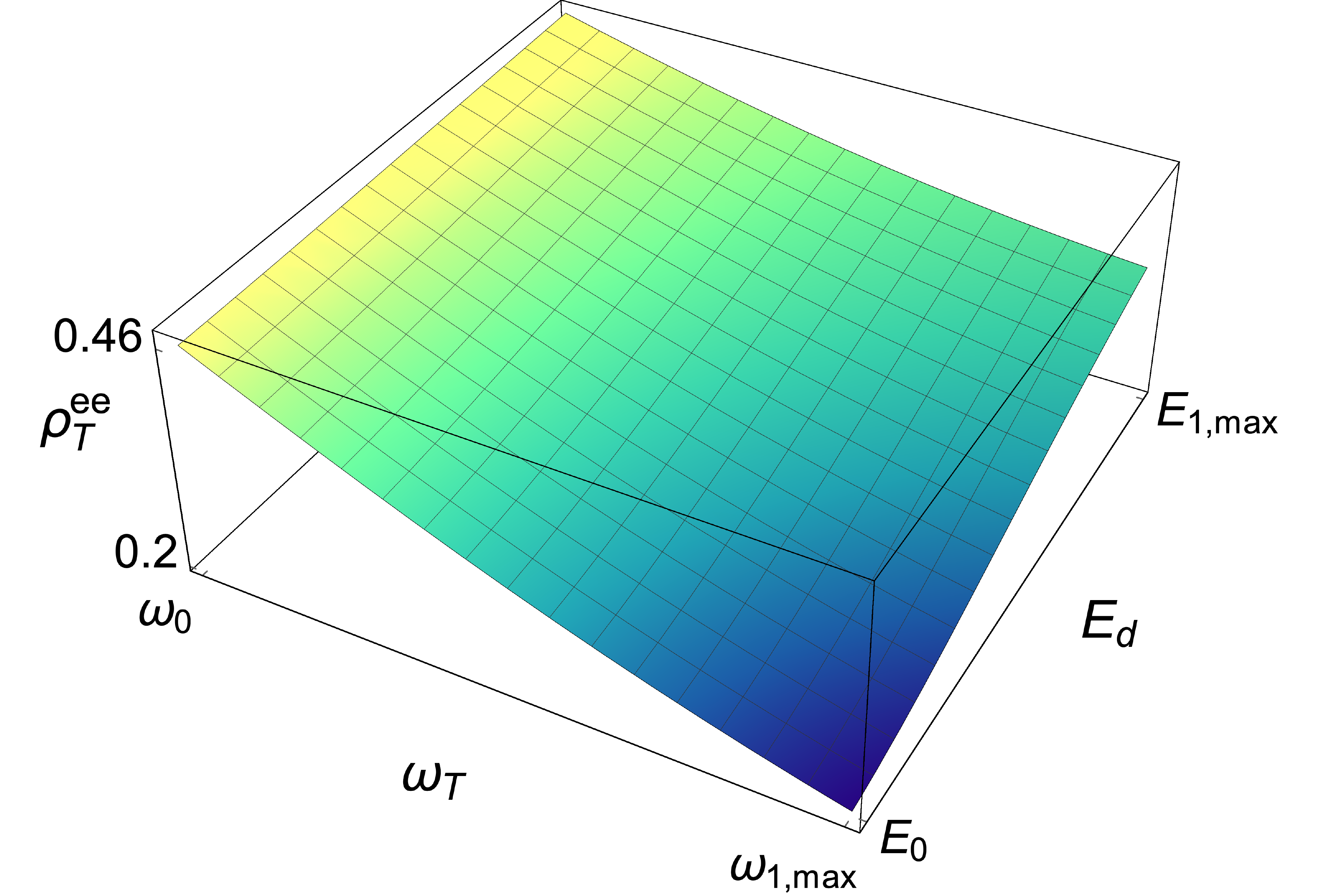}}
\subfigure[Coherence]{\label{fig:d}\includegraphics[width=80mm]{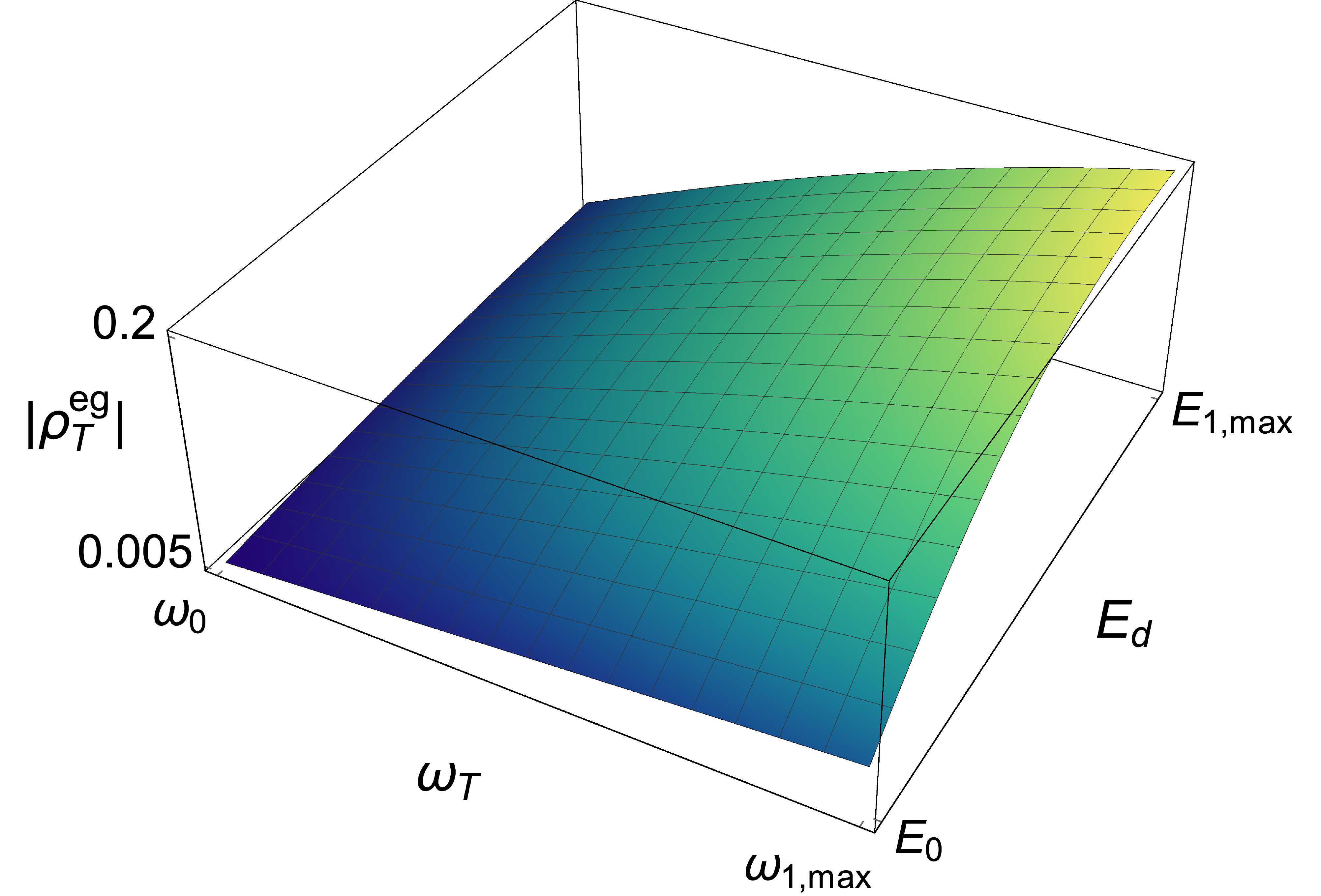}}
\caption{Stationary state's elements $\rho_\mrm{T}^{ee}$ and $|\rho_\mrm{T}^{eg}|$ for different values of $(\omega_\mrm{T},E_d)$. Important amounts of population and quantum coherence changes can be reached during the engine operation.}
\label{state}
\end{figure}

Here, we summarize the explicit expressions of the density matrix elements of $\rho_\mrm{T}^{ss}$ \eqref{stad_state}, which are plotted as a function of $(\omega_{{\rm T}}, E_{\rm d})$ in Fig. \ref{state} . 
\begin{equation}
 \rho_\mrm{T}^{ee}=\frac{\frac{g^2 E^2_d}{\hbar^4\Big[\frac{1}{4}\kappa_\mrm{CPW}^2+(\omega_\mrm{CPW}-\omega)^2\Big]}+\frac{1}{1+e^{\beta\hbar\omega_\mrm{T}}}\Big[\frac{1}{4}\frac{\Gamma^2}{\tanh^2{(\beta\hbar\omega_\mrm{T}}/2)}+(\omega_\mrm{T}-\omega)^2\Big] }{\frac{2g^2 E^2_d}{\hbar^4\Big[\frac{1}{4}\kappa_\mrm{CPW}^2+(\omega_\mrm{CPW}-\omega)^2\Big]}+\Big[\frac{1}{4}\frac{\Gamma^2}{\tanh^2{(\beta\hbar\omega_\mrm{T}/2)}}+(\omega_\mrm{T}-\omega)^2\Big] },
\end{equation}
\begin{equation}
\rho_\mrm{T}^{gg}=\frac{\frac{g^2 E^2_d}{\hbar^4\Big[\frac{1}{4}\kappa_\mrm{CPW}^2+(\omega_\mrm{CPW}-\omega)^2\Big]}+\frac{1}{1+e^{-\beta\hbar\omega_\mrm{T}}}\Big[\frac{1}{4}\frac{\Gamma^2}{\tanh^2{(\beta\hbar\omega_\mrm{T}}/2)}+(\omega_\mrm{T}-\omega)^2\Big] }{\frac{2g^2 E^2_d}{\hbar^4\Big[\frac{1}{4}\kappa_\mrm{CPW}^2+(\omega_\mrm{CPW}-\omega)^2\Big]}+\Big[\frac{1}{4}\frac{\Gamma^2}{\tanh^2{(\beta\hbar\omega_\mrm{T}/2)}}+(\omega_\mrm{T}-\omega)^2\Big] },
\end{equation}
\begin{equation}
\rho_\mrm{T}^{eg}=\frac{\frac{1}{2\hbar}\Big[\frac{\Gamma}{\tanh{(\beta\hbar\omega_\mrm{T}/2)}}i+2(\omega_\mrm{T}-\omega)\Big]\frac{g E_d}{\hbar\Big[i\frac{\kappa_\mrm{CPW}}{2}-(\omega_\mrm{CPW}-\omega)\Big]}}{\frac{2g^2 E^2_d}{\hbar^4\Big[\frac{1}{4}\kappa_\mrm{CPW}^2+(\omega_\mrm{CPW}-\omega)^2\Big]}+\Big[\frac{1}{4}\frac{\Gamma^2}{\tanh^2{(\beta\hbar\omega_\mrm{T}/2)}}+(\omega_\mrm{T}-\omega)^2\Big] }\tanh{(\beta\hbar\omega_\mrm{T}/2)}.
\end{equation}
\newpage
\onecolumngrid
\section{\label{therm_quant}Thermodynamic quantities along each stroke}
In this appendix we summarize the explicit expressions of the thermodynamic quantities $W_i$ and $Q_i$ for $i=1,2,3,4$ and the heat $Q_{+}$ given to the WS. These quantities are obtained by changing quasi-statically the parameters $\omega_\mrm{T}$ and $E_d$ producing a succession of steady states $\hat{\rho}_\mrm{T}^{ss}(\omega_\mrm{T},E_d)$:
\twocolumngrid
\begin{eqnarray}
W_1&=\int_{\omega_0}^{\omega_1}{\tr{\hat{\rho}_\mrm{T}^{ss}(\omega_\mrm{T},E_0)\Big(\frac{\partial \tilde{H}_\mrm{T,RWA}}{\partial\omega_\mrm{T}}\Big)_{E_0}}d\omega_\mrm{T}},\nonumber\\
W_2&=\int_{E_0}^{E_1}{\tr{\hat{\rho}_\mrm{T}^{ss}(\omega_1,E_{d})\Big(\frac{\partial \tilde{H}_\mrm{T,RWA}}{\partial E_d}\Big)_{\omega_1}}d E_d},\nonumber\\
W_3&=\int_{\omega_1}^{\omega_0}{\tr{\hat{\rho}_\mrm{T}^{ss}(\omega_\mrm{T},E_1)\Big(\frac{\partial \tilde{H}_\mrm{T,RWA}}{\partial\omega_\mrm{T}}\Big)_{E_1}}d\omega_\mrm{T}},\nonumber\\
W_4&=\int_{E_1}^{E_0}{\tr{\hat{\rho}_\mrm{T}^{ss}(\omega_0,E_{d})\Big(\frac{\partial \tilde{H}_\mrm{T,RWA}}{\partial E_d}\Big)_{\omega_0}}d E_d}.
\end{eqnarray}
\begin{eqnarray}
Q_{1}&=\int_{\omega_0}^{\omega_1}{\tr{\Big(\frac{\partial \hat{\rho}_\mrm{T}^{ss}}{\partial \omega_\mrm{T}}\Big)_{E_0}\tilde{H}_\mrm{T,RWA}(\omega_\mrm{T},E_0)}d\omega_\mrm{T}},\nonumber\\
Q_{2}&=\int_{E_0}^{E_1}{\tr{\Big(\frac{\partial \hat{\rho}_\mrm{T}^{ss}}{\partial E_d}\Big)_{\omega_1}\tilde{H}_\mrm{T,RWA}(\omega_1,E_d)}d E_d},\nonumber\\
Q_{3}&=\int_{\omega_1}^{\omega_0}{\tr{\Big(\frac{\partial \hat{\rho}_\mrm{T}^{ss}}{\partial \omega_\mrm{T}}\Big)_{E_1}\tilde{H}_\mrm{T,RWA}(\omega_\mrm{T},E_1)}d\omega_\mrm{T}},\nonumber\\
Q_{4}&=\int_{E_1}^{E_0}{\tr{\Big(\frac{\partial \hat{\rho}_\mrm{T}^{ss}}{\partial E_d}\Big)_{\omega_0}\tilde{H}_\mrm{T,RWA}(\omega_0,E_d)}d E_d}.
\end{eqnarray}
\onecolumngrid
\begin{equation}
Q_{+}=\sum_{i=1}^{4}{Q_{+}^{i}}
\end{equation}
with $Q_{+}^{i}$ for $i=1,2,3,4$ given by
\begin{eqnarray}
Q_{+}^{1}&=\int_{\omega_0}^{\omega_1}{\tr{\left(\frac{\partial \hat{\rho}_\mrm{T}^{ss}}{\partial \omega_\mrm{T}}\right)_{E_0}\tilde{H}_\mrm{T,RWA}(\omega_\mrm{T},E_0)}\Theta\left[\tr{\left(\frac{\partial \hat{\rho}_\mrm{T}^{ss}}{\partial \omega_\mrm{T}}\right)_{E_0}\tilde{H}_\mrm{T,RWA}(\omega_\mrm{T},E_0)}d\omega_\mrm{T}\right]},\nonumber\\
Q_{+}^{2}&=\int_{E_0}^{E_1}{\tr{\left(\frac{\partial \hat{\rho}_\mrm{T}^{ss}}{\partial E_d}\right)_{\omega_1}\tilde{H}_\mrm{T,RWA}(\omega_1,E_d)}\Theta\left[\tr{\left(\frac{\partial \hat{\rho}_\mrm{T}^{ss}}{\partial E_d}\right)_{\omega_1}\tilde{H}_\mrm{T,RWA}(\omega_1,E_d)}d E_d\right]},\nonumber\\
Q_{+}^{3}&=\int_{\omega_1}^{\omega_0}{\tr{\left(\frac{\partial \hat{\rho}_\mrm{T}^{ss}}{\partial \omega_\mrm{T}}\right)_{E_1}\tilde{H}_\mrm{T,RWA}(\omega_\mrm{T},E_1)}\Theta\left[\tr{\left(\frac{\partial \hat{\rho}_\mrm{T}^{ss}}{\partial \omega_\mrm{T}}\right)_{E_1}\tilde{H}_\mrm{T,RWA}(\omega_\mrm{T},E_1)}d\omega_\mrm{T}\right]},\nonumber\\
Q_{+}^{4}&=\int_{E_1}^{E_0}{\tr{\left(\frac{\partial \hat{\rho}_\mrm{T}^{ss}}{\partial E_d}\right)_{\omega_0}\tilde{H}_\mrm{T,RWA}(\omega_0,E_d)}\Theta\left[\tr{\left(\frac{\partial \hat{\rho}_\mrm{T}^{ss}}{\partial E_d}\right)_{\omega_0}\tilde{H}_\mrm{T,RWA}(\omega_0,E_d)}d E_d\right]}.
\end{eqnarray}
where the Heaviside function $\Theta[\cdot]$ is inside the integral, selecting only the positive contributions (heat given to the WS) along the stroke.
\twocolumngrid
\bibliography{apssamp}

\end{document}